\documentclass[%
 preprint,
superscriptaddress,
showpacs,
 amsmath,amssymb,
 aps,
 pre,
]{revtex4-1}

\usepackage{graphicx}% Include figure files
\usepackage{dcolumn}% Align table columns on decimal point
\usepackage{bm}% bold math
\usepackage{natbib}
\usepackage{color}

\usepackage[english]{babel}
\usepackage[utf8]{inputenc}

\begin{document}

\title{
Characterizing complex networks using Entropy-degree
diagrams: \\ unveiling changes in functional brain connectivity \\ induced by Ayahuasca
}

% Force line breaks with \\
%\thanks{}%
\author{A. Viol}
 \email{aline.viol@bccn-berlin.de}
 \affiliation {Institute of Theoretical Physics, Technische Universit\"at, Berlin, Hardenbergstra\ss{}e 36, 10623 Berlin, Germany}
\affiliation{Bernstein Center for Computational Neuroscience Berlin, Humboldt-Universit{\"a}t zu Berlin, Philippstra{\ss}e 13, 10115 Berlin, Germany}

\author{Fernanda Palhano-Fontes}

\affiliation {Brain Institute, Universidade Federal do Rio Grande do Norte, 59078-970 Natal--RN, Brazil}

\author{Heloisa Onias}

\affiliation {Brain Institute, Universidade Federal do Rio Grande do Norte, 59078-970 Natal--RN, Brazil}

\author{Draulio B. de Araujo}%

\affiliation {Brain Institute, Universidade Federal do Rio Grande do Norte, 59078-970 Natal--RN, Brazil}

 \author{Philipp H\"ovel}
  \affiliation {School of Mathematical Sciences, University College Cork, Western Road, Cork, Ireland}
 \affiliation{Bernstein Center for Computational Neuroscience Berlin, Humboldt-Universit{\"a}t zu Berlin, Philippstra{\ss}e 13, 10115 Berlin, Germany}

\author{G. M. Viswanathan}%
\email{gandhi@fisica.ufrn.br}
\affiliation {{Department of Physics, Universidade Federal do Rio Grande do Norte,} 59078-970 Natal--RN, Brazil}
\affiliation {  National Institute of Science and Technology of Complex Systems 
Universidade Federal do Rio Grande do Norte, 59078-970 Natal--RN, Brazil}

\date{\today}% It is always \today, today,
\begin{abstract}
Open problems abound in the theory of complex networks, which has
found successful application to diverse fields of science.
With the aim of further advancing the understanding of the brain's
functional connectivity, we propose to evaluate a network metric which
we term the {\it geodesic entropy}. This entropy, in a way that can be
made precise, quantifies the Shannon entropy of the distance
distribution to a specific node from all other nodes.
Measurements of geodesic entropy allow for the characterization of the
structural information of a network that takes into account the
distinct role of each node into the network topology.
The measurement and characterization of this structural information
has the potential to greatly improve our understanding of sustained
activity and other emergent behaviors in networks, such as
self-organized criticality sometimes seen in such contexts.
We apply these concepts and methods to study the effects of how the
psychedelic Ayahuasca affects the functional connectivity of the human
brain.
We show that the geodesic entropy is able to differentiate the
functional networks of the human brain in two different states of
consciousness in the resting state: (i) the ordinary waking state and
(ii) a state altered by ingestion of the Ayahuasca.
% --a brew containing the psychedelic substance N, N-Dimethyltryptamine (DMT).   
%
The entropy of the nodes of brain networks from subjects under the
influence of Ayahuasca diverge significantly from those of the
ordinary waking state.  The functional brain networks from subjects in
the altered state have, on average, a larger geodesic entropy compared
to the ordinary state.
We conclude that geodesic entropy is a useful tool for analyzing complex
networks and discuss how and why it may bring even further valuable
insights into the study of the human brain and other empirical networks.
\end{abstract}

%\pacs{64.60.aq, %Networks
%87.19.lf, %-Neuroscience MRI
%89.75.Fb, % Structures and organization in complex systems
%87.18.-h, %Biological complexity
%89.70.Cf, %Entropy and other measures of information
%87.61.-c, %Magnetic resonance imaging
%  87.19.lo, %Information theory
 %  02.10.Ox%	Combinatorics; graph theory

\maketitle

%\tableofcontents
%
\section{Introduction}
In the last few decades, new scientific fields have taken advantages
of complex network approaches.
This interest emerged, in part, by virtue of technological advances
that generate new datasets in computational, social, biological, among
others sciences.
Examples include modern brain mapping techniques, such as functional
magnetic resonance imaging (fMRI), that have provided previously
inaccessible information about interaction patterns in the human brain
\cite{HAY06}.
The theory of complex networks has proven to be a crucial tool to
understand the  interactions and dynamics in large systems.

Attempts to characterize those new datasets bring up the challenge of extracting
relevant features regarding the network's structure.
One of the main concerns is to identify the role of each node in the network and how 
the nodes cooperate to give rise to 
 emergent behaviors.  
The majority of measurements that have been proposed in the last few
decades allow the ranking of nodes' importance by the number of
connections, centrality, etc.  \cite{NEW05a,HEU13,HOU12}.

Instead of ranking a node's relative importance, we ask how the nodes
contribute locally to the global connectivity of the network, with the
aim of better understanding the individualize role played by each node
in the network.  We quantitatively describe these roles, as well as
the structural information of the diversity of interactions between
nodes.
The nodes in a network interact with their neighbors and, indirectly,
with the neighbors of neighbors; and also with more distant nodes with
even greater ``neighborhood radius'' (Figure \ref{scheme}).

We aim to quantify the diversity of influences on a given node, of all
other nodes over the whole network.  For each node, we calculate the
Shannon entropy functional \cite{SHAly} of the probability distribution
of the geodesic distances between each node and all other nodes. We
call this measurement {\it geodesic entropy}.
Nodes with a great diversity of influences (i.e., with high geodesic
entropy) may play an important role in, for example, to guarantee
specialization of functional patterns. 
Besides, nodes with a low diversity of influences may guarantee
constraints relevant to network robustness.
The ``fine tuning'' of the distribution of distances, quantified by
the Shannon entropy, may be a key to understanding how emergent
behaviors arize.

We illustrate and apply our method to real network data.  We use the
geodesic entropy to analyze human brain functional networks under the
influence of the psychedelic Ayahuasca -- a brew from the Amazonian
indigenous cultures that contains the serotonergic psychedelic N,
N-Dimethyltryptamine (DMT) and monoamine oxidase inhibitors (MAOi)
\cite{RIB03a}.

Ayahuasca ingestion may cause deep changes in the cognition and
perceptions, promoting substantial alterations in the sense of the
reality and the self \cite{SHA02d,RIB01}.
According to the neural correlate hypothesis, we expect to find features on functional brain
networks that can be correlated to this specific consciousness state. 
We evaluate the networks extracted from fMRI data
acquired from the same group of subjects in two sections: before and 40 minutes after Ayahuasca intake. 
The geodesic entropy is able to identify a specific behavior for networks related to the psychedelic
state of consciousness: the nodes of functional brain networks under Ayahuasca effects tend to have a greater geodesic entropy than the ordinary condition.  
\section{Methods}
A complex network is a schematic representation of the relations (links) between elements (nodes)  
of a system with a nontrivial topology of interactions \cite{NEW10,ALB02a}. 
Consider a non-weighted undirected network $G(\nu,\xi)$,  where $\nu$ is a set with $N$ nodes 
and $\xi$ is the set of links. It is represented numerically
by a $N \times N$ adjacency matrix $A_{i,j}$: if a pair of nodes $i$ and $j$ are connected,
the matrix element is $A_{i,j}=1$ and $A_{i,j}=0$ otherwise. % \cite{NEW10}.
The nodes are connected if the elements that they represent share some kind of information or have mutual influences. 
The number of links that have each node is termed degree. % \cite{NEW10}.
The statistics of the degrees in a network is quantified by the degree distribution,
a histogram of degrees considering the whole network \cite{NEW10}.   

Nodes directly connected are called first neighbors.
A node can also influence and be influenced by the neighbors of its neighbors, called second neighbors.
Considering a connected network, the influences may be extended to all neighborhood radius. 
Our goal is to quantify the amount of information involved in the diversity of influence
extending over the network.
For this purpose, we calculated the Shannon entropy \cite{SHAly}
considering the statistics of distances between a node and all their
neighborhood radius.

Distances in network theory are related to the paths lengths. 
By definition, a path length $\Gamma_{i,j}$ is the number of consecutive links between the pair of nodes $i$ and $j$, following a specific trail. 
The shortest path length ($D(i,j)=min(\{ \Gamma_{i,j}\}) $) defines geodesic distance between two nodes \cite{RUB10a}.
 The geodesic distance has been used in several network characterizations such as small-world networks \cite{WAT98a}. 

By looking at the distribution of geodesic distances for a given node,
we can better understand the role played in the network by that
particular node.
Quantifying the diversity of influences due the geodesic distances brings to light the rules of how
the information is distributed in the network. 

\begin{figure}[ht]
%\centering
\hspace{-0.6cm}
\includegraphics[width=0.55\linewidth]{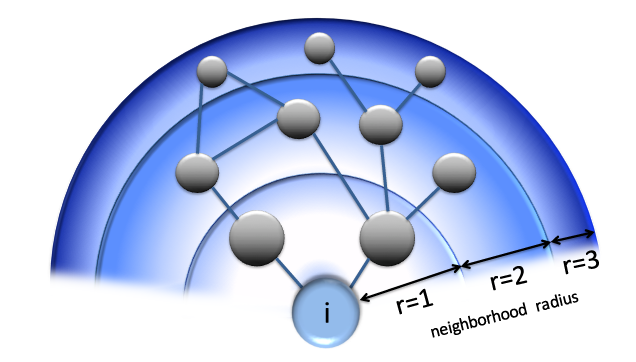}  
\includegraphics[width=0.47\linewidth]{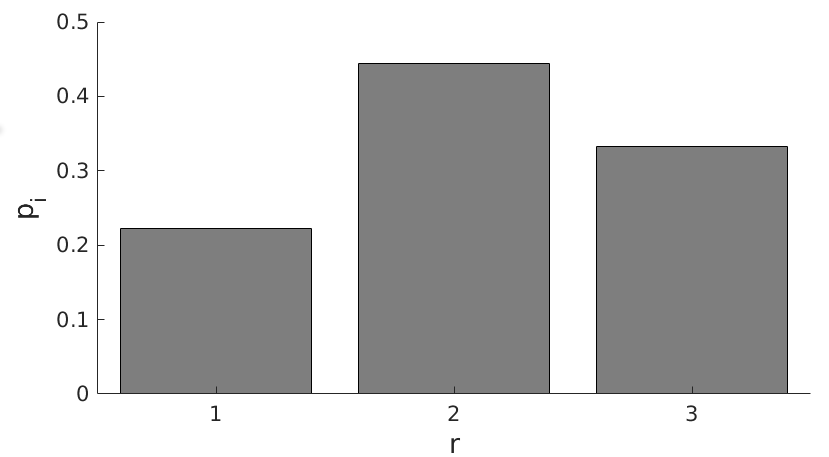}  
\caption{Schematic representation of neighborhood radius $r$  and its probability distribution.
The  left panel  shows three neighborhood radius for the node i. 
The nodes within the neighborhood $r=1$, $r=2$, $r=3$ are, respectively,
1, 2 and 3 links distant from the node $i$. On the right is the probability distribution of the geodesic distances for this network.}
\label{scheme}
\end{figure}

We define $P_i\{p_i(r), 1\leq r\leq max(D(i,\{j\}\}$ 
as a probability mass function of find a node in the neighborhood ratio $r$ of the node
$i$. That is, the probability of, in a random
choose, one selects a node $j$ from the set of the remain nodes
($\{j\} := j \in \nu / ~j\neq i$) with the geodesic distance
$D(i,j)=r$.  This probability is defined as:

\begin{equation}
p_{i} (r)= \frac{1}{(N-1)}\sum_{\{j\}}{ \delta_{D(i,j),r} } ~;
\end{equation}
where neighborhood radius $r$ assumes values according to the interval ($1\leq r\leq max(D(i,\{j\}$). 
See an illustration in Figure \ref{scheme}.

The distribution $P_i(r) $ contains
  information about the connectivity across multiple
  links of a network.  For illustration, consider hypercubic lattices of dimension
  $D$ with links only between neighboring nodes. The distribution
  $P_i(r)$ scales according to $P_i(r) \sim r^{D-1}$, because nodes a
  fixed distance $r$ away lie on the (hyper)surface of constant
  distance to the node $i$, where in $D$ dimensions, this surface has
  dimension $D-1$.
  Hence, it makes sense that the characerization of the distribution
  $P_i(r)$ has the potential to provide insights into network
  connectivity.

The geodesic entropy is given by:
  \begin{equation}  
  s^{{g}}_i[P_i]= -\sum_{r=1}^{r_{max}} p_i(r)  \log{p_i(r)}~; 
  \end{equation}
where $r_{max}=max(D(i,\{j\})$. The value of $s^{{g}}_i$ does not depend on the network size for greater networks ($N \gg r_{max})$.
The characteristic geodesic entropy of a network is defined by:
\begin{equation}
{S}^{{g}}= \frac{1}{N}\sum^{N}_{i=1} s^{{g}}_i ~;
\end{equation}

Distinct from the entropy of the degree distribution, that quantifies
the constraints imposed by the network degree distribution \cite{VIO17a},
the geodesic entropy quantifies the information due to the intrinsic configuration of network structure. 
Networks with different structures can share the same degree distribution, that is, they can be degenerate
in the entropy of the degree distribution.
The characteristic geodesic entropy can lift the degeneracy of those networks. 
%discriminate those networks. 
%
Besides, the geodesic entropy is a measurement more appropriate to characterize the nodes role and the underlying trends in the network topology. 

We briefly compare and relate the geodesic entropy to similar quantities that have been used to study
networks. The use of geodesic distances to evaluate Shannon entropy was firstly proposed by Chen and collaborators \cite{CHE14}.
Instead define the entropy per node, they defined a global entropy ($I_r(G)$) considering only one specific value $r$ of geodesic distance.
A recent work from  Stella and Domenico proposes a similar formula proposed in this work to characterize centrality by mean of Shannon entropy \cite{STE18}.
Their proposes differente from ours by a normalization factor that depends on 
$r_{max}$. It limitates the entropy to be defined between $0$ and $1$. 
This normalization does not take in consideration the increase on entropy due the increase of maximum radius $r_{max}$.
In contrast to the above methods, the geodesic entropy we propose here allows the evaluate the influence of the maximum neighborhood radius, as well as its dependence of network size, and to depict the role of each node in the network. 

\subsection{Entropy-degree diagram}
\begin{figure}[ht]
\hspace{-0.1cm}\includegraphics[width=0.35\linewidth]{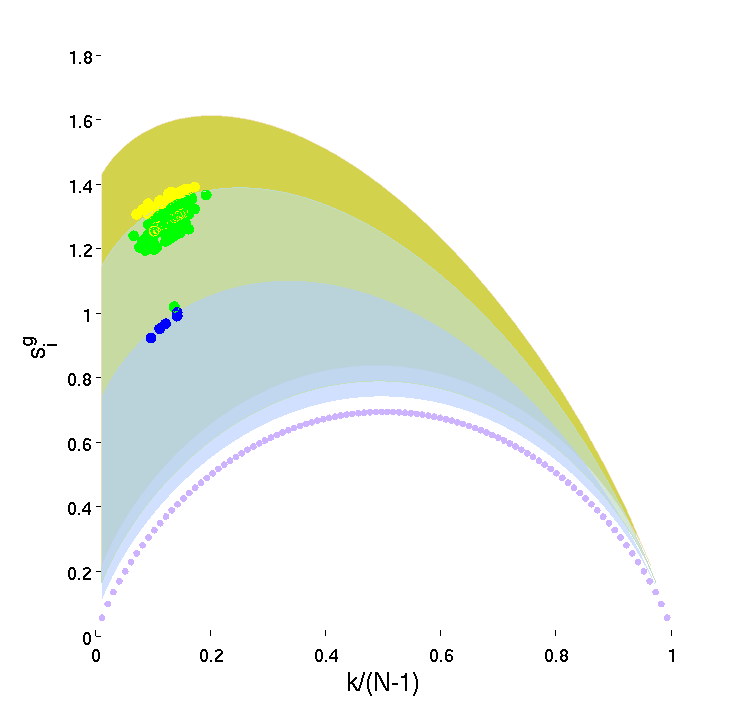} 
\includegraphics[width=0.37\linewidth]{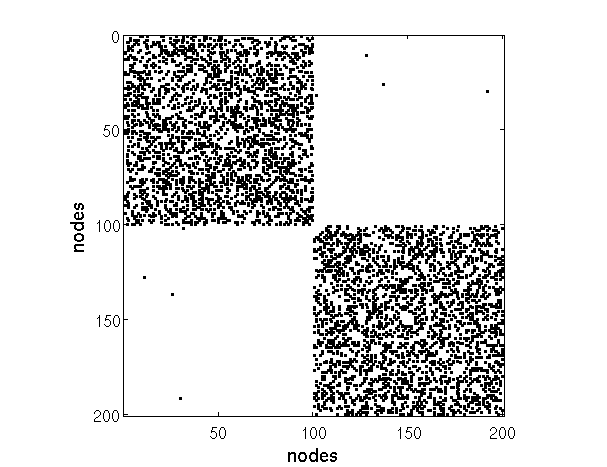}

\hspace{-0.1cm}\includegraphics[width=0.35\linewidth]{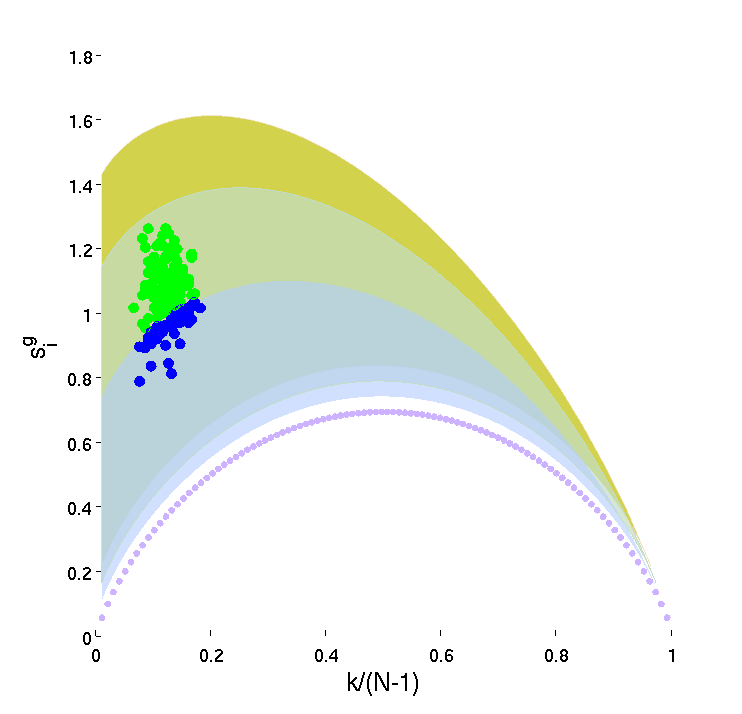} 
\includegraphics[width=0.37\linewidth]{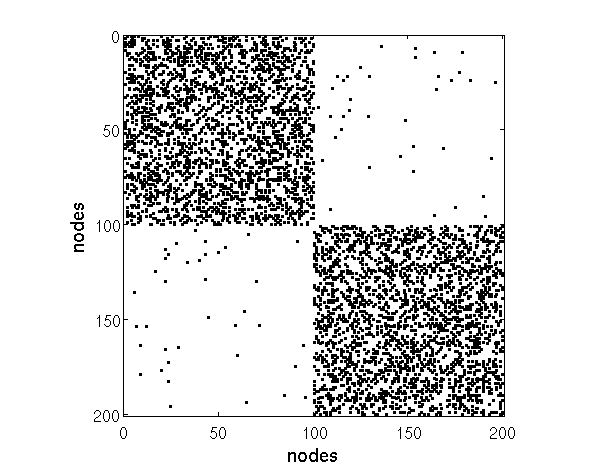}

\hspace{-0.1cm}\includegraphics[width=0.35\linewidth]{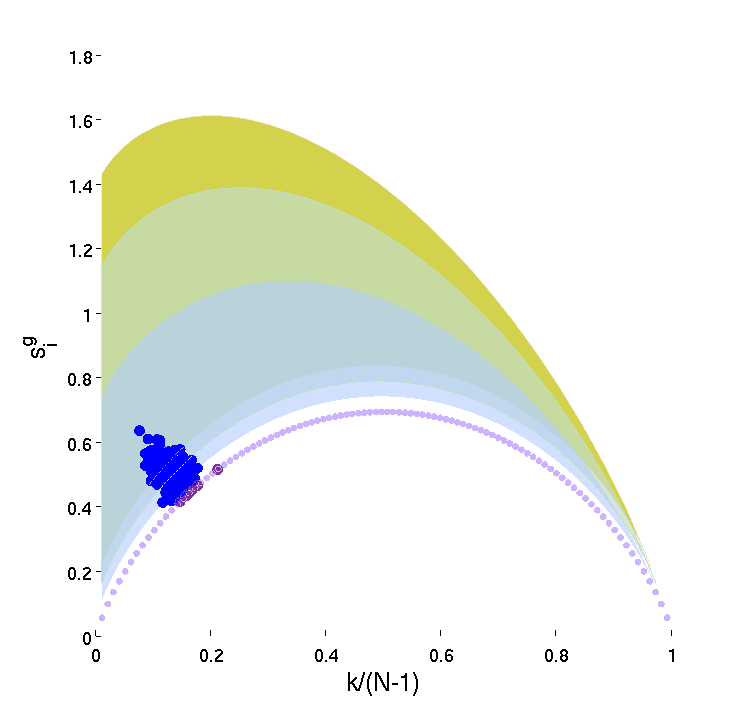} 
\includegraphics[width=0.37\linewidth]{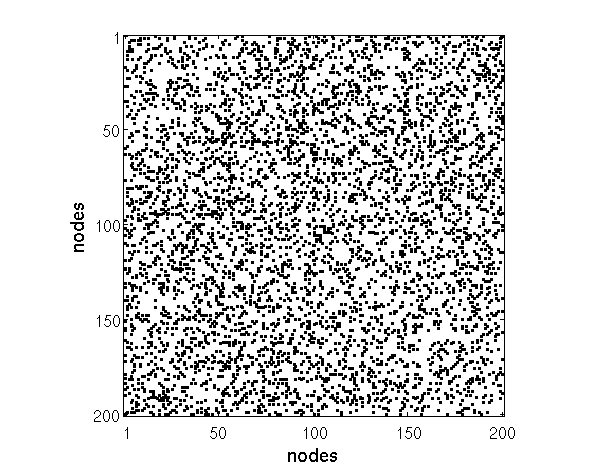} 

\caption{
Illustration of entropy-degree diagram for three different artificially generated networks with the same number of nodes, 
links, same degree distribution ($\langle k \rangle=25$), but different structural configurations.
On the right side of each entropy-degree diagram is the adjacency matrix of the corresponding network.The characteristic geodesic entropy are $S^{{g}}=1.26 ~ \mbox{nats}$, $S^{{g}}=0.98 ~ \mbox{nats}$, $S^{{g}}=0.52 ~ \mbox{nats}$ 
reespectvely from panels up to down. 
The colors purple, blue, green and red are defined according to the maximum neighborhood radius $r_{max}=2,3,4$ and $5$. 
The watermark regions delimit the space of possibilities for each value of the maximum neighborhood.
The minimum entropies possible are delimited by the purple curve and depends on the node degree.
}
\label{skdiagram}
\end{figure}

\begin{figure}[h!]
\hspace{-0.1cm}\includegraphics[width=0.48\linewidth]{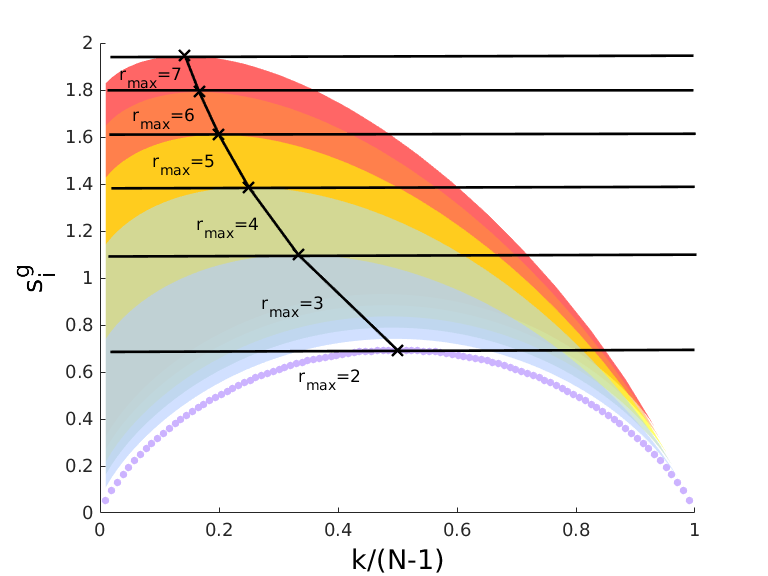} 
\includegraphics[width=0.48\linewidth]{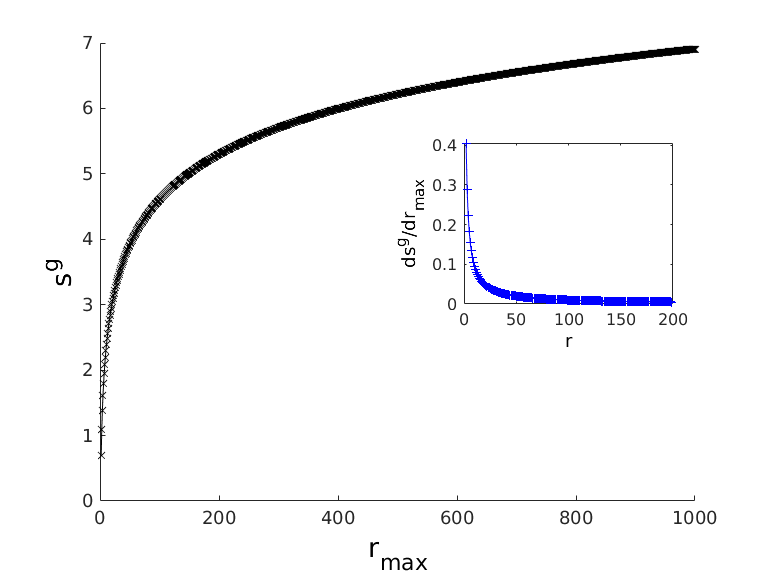}  
 \includegraphics[width=0.48\linewidth]{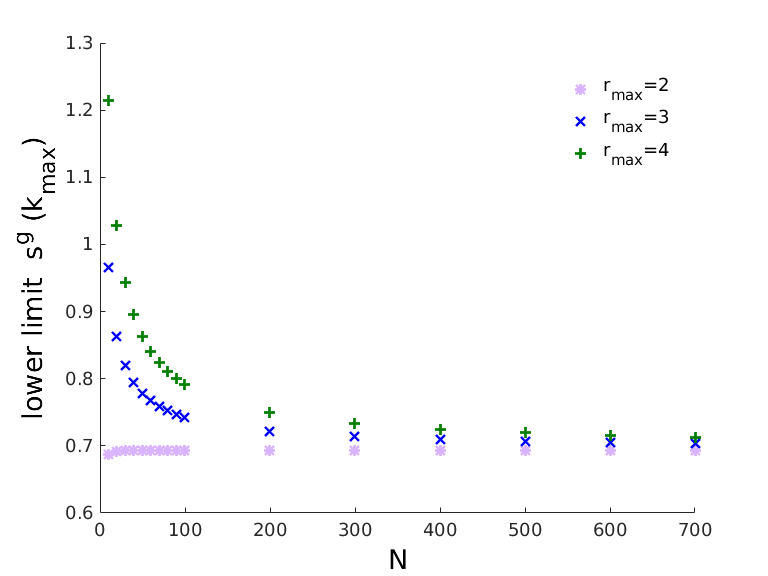}  

\caption{Universal relations for geodesic entropy. In the sk-diagram on left, 
the colored regions delimit the regions of nodes with maximum neighborhood values ($r_{max}$) indicated by the labels.
% $r_{max}=2$ (purple), $r_{max}=3$ (blue),$r_{max}=4$(green),$r_{max}=5$(yellow),$r_{max}=6$ (orange),$r_{max}=7$ (red).  
The maximum value of each region is in $s^{ {g}}_i \approx ln(r_{max})$, that correspond to  $ k \approx 1 \backslash r_{max}$.
The middle plot shows the variance in the maximum entropy due the increase of neighborhood radius $r_{max}$. 
Its increases are inversely proportional to $r_{max}$ ($\Delta s^{ {g}}_{max} \approx r_{max}^{-1} \Delta  r_{max}$).
Note that for small values of $r_{max}$, its increasing will result in an increase in contribution
to the entropy. Nevertheless, for large $r_{max}$ the contribution does not change significantly.
Note that none of these values depend on network size. The finite size effect appears in the lower limit. 
For large networks, all regions will be delimited within the fist curve ($r_{max}=2$). The lower limit will depend on the network size for finite networks. 
The right plot shows the influence of the network size in the lower limits.  
}
\label{sklevels}
\end{figure}

We introduce here the entropy-degree diagram, a viewer tool to help to map the role of nodes into the network. 
Entropy-degree diagram is built plotting the geodesic entropy ($s^{{g}}_{i}$) {\it versus} 
the nodal degree $k$ normalized by the maximum number of connections possible ($k/(N-1)$)
for all nodes belonging to the network. This normalization allows we compare networks with different sizes.
Figure \ref{skdiagram} shows the entropy-degree diagram for 3 networks that
share the same number of nodes and links, have the same degree distribution but have different structures. 
Each marker ($\bullet$) represents a node. 
We used here colors as a didactic artifact to improve the visualization (it can be neglected to build the entropy-degree diagram). 
The colors are defined according to their maximum neighborhood radius ($r_{max}$), that is, 
the greatest geodesic distance between the given node and the remaining nodes.
The watermark regions follow the same colors and delimit the space of possibilities for each value of $r_{max}$.
For example, the purple curve delimits the possible positions on the diagram for nodes with first and second neighbors.
The region in blue delimits the positions for nodes with first, second and third neighbors and it follows for the others regions.
The up limit of each $r_{max}$ region are peaked at ($k\approx 1/r_{max}$, $ s^{{g}}\approx\ln r_{max}$). 
Note the values have no dependence with the network size. They depend only on the network structure. 
The magnitude of the increment in the geodesic entropy due to the increase of $r_{max}$ 
is inversely proportional to $r_{max}$ , ($\Delta s^{{g}} \approx r_{max}^{-1} \Delta r_{max})$. 
That means there is a limit in which the increase of maximum geodesic distances (increase the sparsity) contributes
significantly to the network entropy. 
The lower limits will be affected by the size of the network and converge to the first curve ($r_{max}=2$) 
for large networks. See Figure  \ref{sklevels}. 
We would like to let open the question if it could explain some optimization patterns in real networks. 

The entropy-degree diagram helps to visualize how the information is distributed across the network.  
The nodes with high entropy comprise more information. 
Their interactions into the networks are more ``flexible". 
That is, they are in a position where the diversity of interactions
is arranged in a way that allows holding more information. 
The opposite can be affirmed to nodes with low entropy. 
\vspace{-0.5cm}

\subsection{Geodesic entropy of functional brain networks under Ayahuasca influence}
We use the geodesic entropy to evaluate functional brain networks in different states of consciousness:
ordinary state and psychedelic state induced by Ayahuasca. 
Ayahuasca is a sacred brew from Amazonian indigenous culture made with two plants from Amazonian flora -- 
the leaves of the bush Psychotria Viridis, that contains N, N-Dimethyltryptamine (DMT), 
and the vine Banisteriopsis caapi, that contains monoamine oxidase inhibitors MAOi \cite{RIB03a}.
The DMT is a serotonergic psychedelic similar to LSD \cite{HOF13a,TOR18}, and mescaline but fast metabolized by the human body.
The MAOi's act slowing down this degradation, allowing the DMT to cross the blood-brain barrier and
enabling hours of psychedelic experience \cite{RIB03a}. 
For more information about Ayahuasca we referee \cite{shanon2002antipodes,labate2014ayahuasca,labate2014prohibition,labate2013therapeutic,riba2001}.

\subsubsection{Data}
The experimental procedures were performed in accordance with the guidelines and regulations
approved by the Ethics and Research Committee of the University of S\~ao Paulo at Ribeir\~ao Preto
(process number 14672/2006). All volunteers sign a written informed consent.
The fMRI data were acquired from 10 healthy adult volunteers (mean age 31.3, from 24 to 47 years, 5 women)
with no history of neurological or psychiatric disorders -- evaluated by DSM-IV structured interview \cite{american2000diagnostic}.
They have at least 8 years of formal educational and minimum Ayahuasca use time of 5 years.
They were in absence of any medication for at least 3 months prior to the acquisition
and also had not take nicotine, caffeine, and alcohol prior to the acquisition.
Each volunteer ingested about 120-200 mL (2.2 mL/kg of body weight) of Ayahuasca.
The chromatography analysis detected on the brew 0.8 mg/mL of DMT, 0.21 mg/mL of harmine
and no harmaline at the threshold of 0.02 mg/mL \cite{draulio2012}.
The volunteers were submitted to two sections of fMRI scanning: one before
and other 40 minutes after Ayahuasca intake when the subjective effects can be observed.
In both cases, volunteers were requested to be in an awake resting state, that is
lying with their eyes closed, without performing any task.
The samples of one volunteer were excluded from the dataset due to excessive head movement.
\subsubsection{Obtaining functional networks from fMRI data}

The methods to extract the networks from the fMRI data used here are the same performed in the reference \cite{VIO17a}.
The pre-processing of fMRI data was made according to standard guidelines.
We performed spatial smoothing (Gaussian kernel, FWHM = 5 mm) and correction of slice-timing and head motion. 
We evaluated 9 regressors using a General Linear Model (GLM): 6 regressors to movement correction, 1 to white matter
signal, 1 to cerebrospinal fluid and 1 to global signal
\footnote{We used FSL Software, 
a free library of statistical tools available by Oxford Centre
for Functional MRI of the Brain ({http://www.ndcn.ox.ac.uk/divisions/fmrib})}.
The images were spatially normalized according to the Montreal Neurologic Institute
(MNI152 template) anatomical standard space using a linear transformation. 
We evaluated the band-pass filter using maximum overlap wavelet transform (MODWT), 
considering the Daubechies wavelet to split the signal into 4 scales
of distinct frequency bands. 
We choose the scale 3 (frequency band $\approx 0.03-0.07$ Hz) to be in agreement 
with the literature that considers the low frequency ($\approx 0.01$ to $0.1$ Hz), 
preeminent on resting states \cite{Fransson2005}.

We parcellate each image into 110 cortical anatomical regions according
to the Harvard-Oxford cortical and subcortical structural atlas
(threshold of $> 25\%$, using FMRIB software, an FSL library).
We evaluated only 104 cortical regions because of an acquisition limitations for some subjects.
The cortical regions were used to define the nodes of the brain networks and the correlation between their signals to define the links. 
The signals corresponding to each cortical region were obtained averaging the time series of all voxels (3D regular grid) into them
(using Marsbar, SPM toolbox). 
We calculate the Pearson correlation of temporal series of all possible pairs of cortical regions, 
yielding a cross-correlation matrix.
Thus, we have for each sample (before and after Ayahuasca of all subjects) 
a 104$\times$104 correlation matrix considered as an estimative of the brain functional connectivity.
Since the cortical regions define the nodes, the correlation matrices were used to define the links of the functional brain networks.  

For each sample, we generated a set of symmetric binary adjacency matrices by
thresholding the absolute value of their correlation matrices.
Precisely, whether the absolute value of the element matrix is larger
than the defined threshold, a link is formed ($A_{i,j}=1$), otherwise, no link is formed
($A_{i,j}=0$).
We choose a range of thresholds that ensure the networks were fully connected but also sparse.
We adopted the same criteria of references \cite{onias2014,schoter2012, liuyong2008,VIO17a}.
We consider the network with lower global efficiency and greater local efficiency than its
randomized version \cite{Maslov2002}.
We fixed the same band of thresholds for all samples, allowing a more accurate comparison. 
It was necessary to exclude two subjects from our analysis due to a trade-off in the range,
leaving 7 subjects (4 women).
As long as we intend to evaluate the difference between topological features of networks before and after Ayahuasca
intake, we compare networks with the same density of links.
The chosen threshold correlation range is $0.28 \leq \eta \leq 0.37$ that yield networks
with mean degree in the range $24 \leq \langle k\rangle \leq 39$.
Summarizing, we created two sets of networks (before and after Ayahuasca intake)
that allow 16 different comparisons (i.e. of differing mean degrees) for each subject's sample. 
The reader can find further details in the reference \cite{VIO17a}.

\section{Results}
\begin{figure}[ht!]
\hspace{-0.25cm}\includegraphics[width=0.03\linewidth]{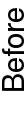} 
\includegraphics[width=0.32\linewidth]{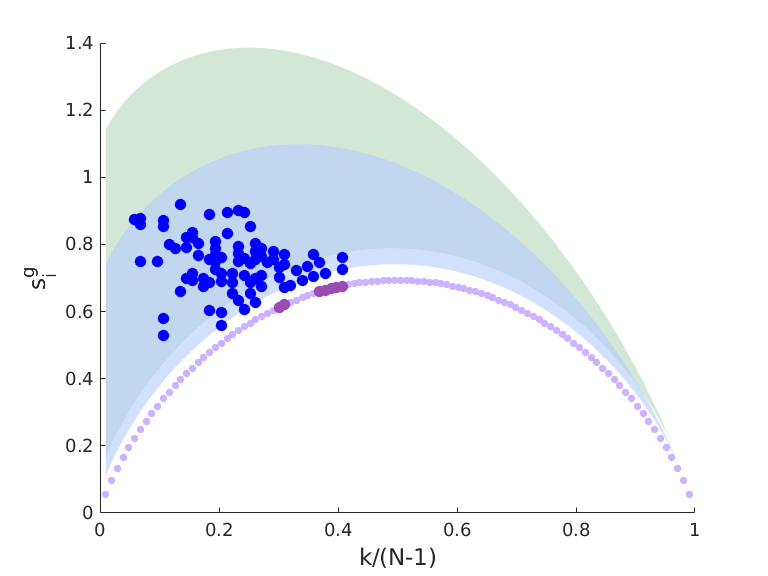} 
\includegraphics[width=0.32\linewidth]{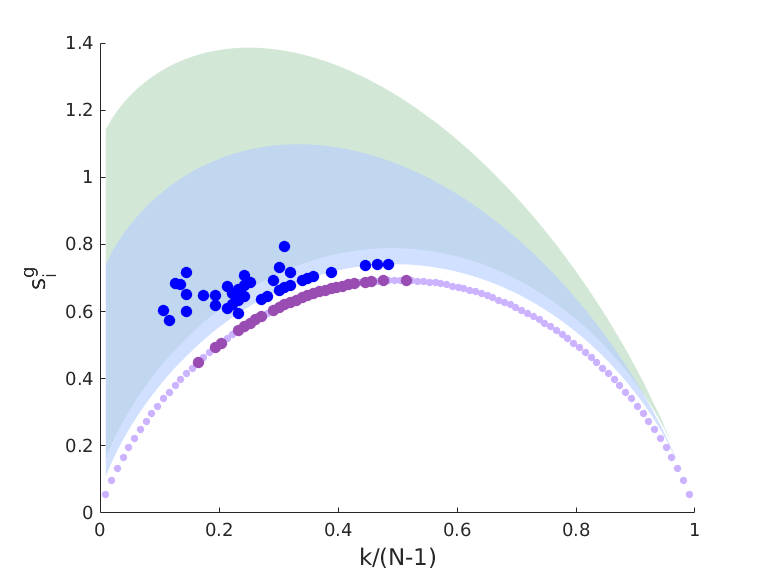}

\hspace{-0.25cm}\includegraphics[width=0.03\linewidth]{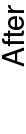} 
\includegraphics[width=0.32\linewidth]{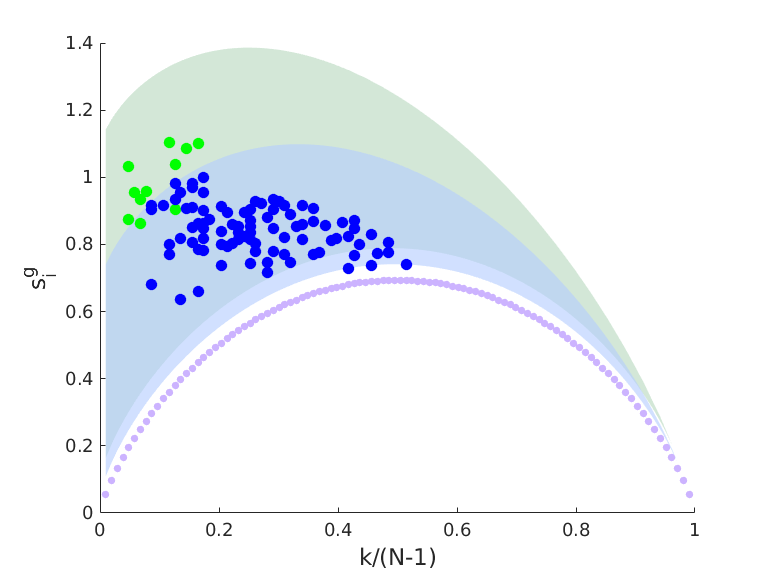} 
\includegraphics[width=0.32\linewidth]{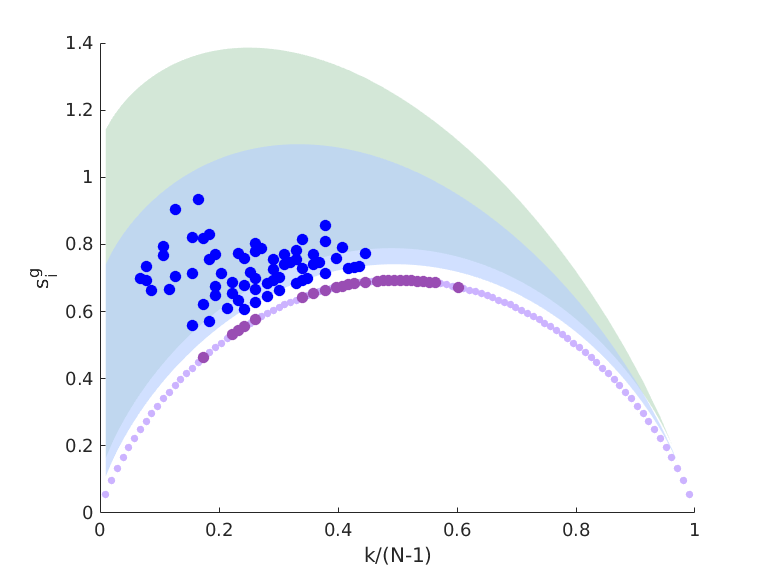} 
\caption{entropy-degree diagram before and after Ayahuasca. 
The panels depict the entropy-degree diagrams for one of the subjects before 
(upper row) and after (bottom row) Ayahuasca intake for networks with mean degree
$\langle k \rangle=25$ and $\langle k \rangle=32$ respectively. 
The colors follow the same rules of figure \ref{skdiagram}.
Note the nodes after Ayahuasca tends to occupy populate regions in the diagram of higher entropy.}

\label{dg}
\end{figure}

\begin{figure}[ht]
    
  \includegraphics[width=\linewidth]{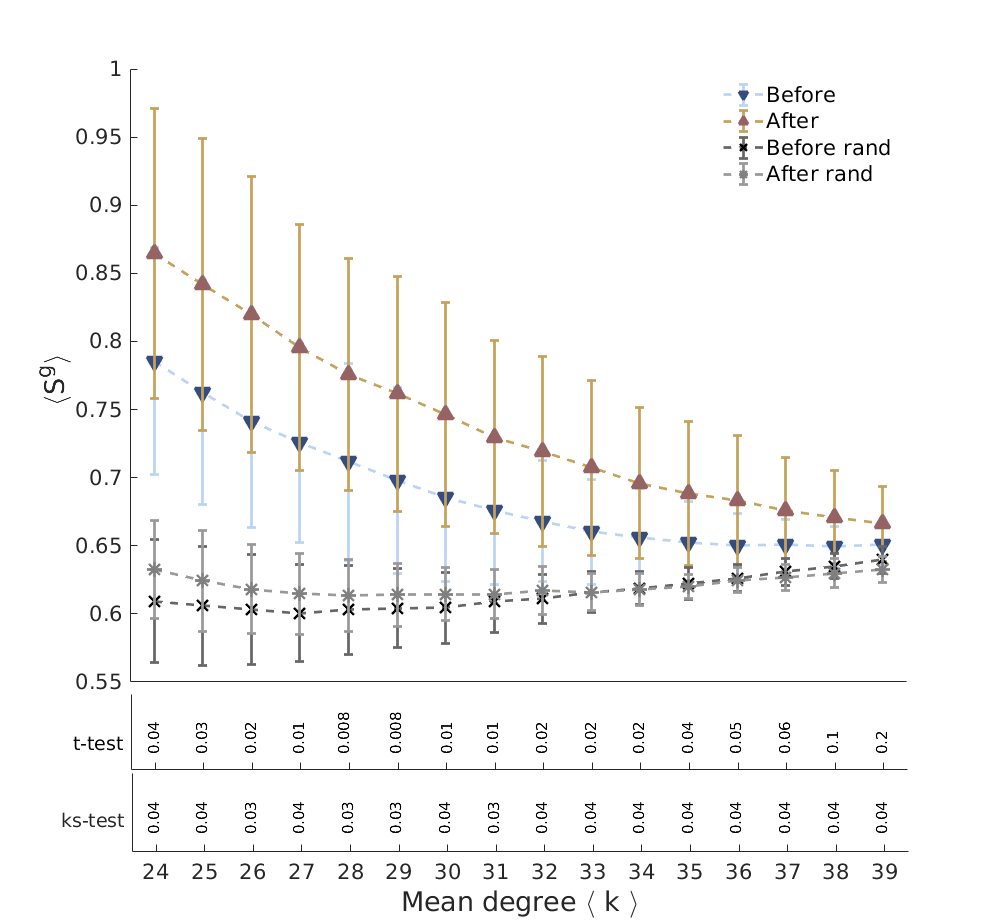}
   \caption{Mean geodesic entropy before and after Ayahuasca. 
The graphic shows the mean geodesic entropy under all subjects for before and after Ayahuasca 
for networks with different densities (mean the degree from 
$<k>=24$ to $<k>=39$ ). The mean geodesic entropy is greater for all networks densities. }
  \label{meanentropy}
  \end{figure}

Figure \ref{dg} shows the entropy-degree diagram of one of the subjects before and after Ayahuasca intake 
for networks with mean degree $\langle k \rangle=25$ and $\langle k \rangle=32$.
Note that the nodes in the entropy-degree diagram after Ayahuasca tend to have higher entropy. 
All subjects presented similar behavior. See supplementary material.
Figure \ref{boxplot} shows the divergences of the characteristic geodesic entropies between
after and before Ayahuasca  for each subject by comparing pair of networks with the same density of links. 
The boxplot depicts the distribution of characteristic geodesic entropy differences 
($\Delta S^{\mbox{g}}=S_{after}^{{g}}-S_{before}^{{g}}$) of networks with the same mean degree. 
Note the characteristic geodesic entropy increases for all subjects after Ayahuasca intake. 
Figure \ref{meanentropy} shows the contrast of the characteristic geodesic entropy of networks  
with the same mean degree (same densities of links) averaged over all subjects before
(blue) and after (brown) Ayahuasca intake.
The increasing also appear in this graphic suggesting that characteristic geodesic entropy
of functional networks under Ayahuasca influence tends to be higher than in ordinary condition.

The black and gray curves show the characteristic geodesic entropy for the randomized versions of
the networks before and after Ayahuasca respectively. 
We used the Maslov algorithm \cite{Maslov2002} to randomize the links of networks keeping
their degree distribution unchanged. 
In other words, the Maslov randomization breaks all structural trends that do not depend on the degree distribution. 
Note that the randomization reduces the entropy in both conditions and no
considerable divergence was found between the randomized curves. 
These results mean that the change in geodesic entropy we detected
before and after Ayahuasca intake is related to underlying trends of
the network structure. They do not result from the known changes in
degree distribution \cite{VIO17a}.

\begin{figure}[ht]
    \includegraphics[width=\linewidth]{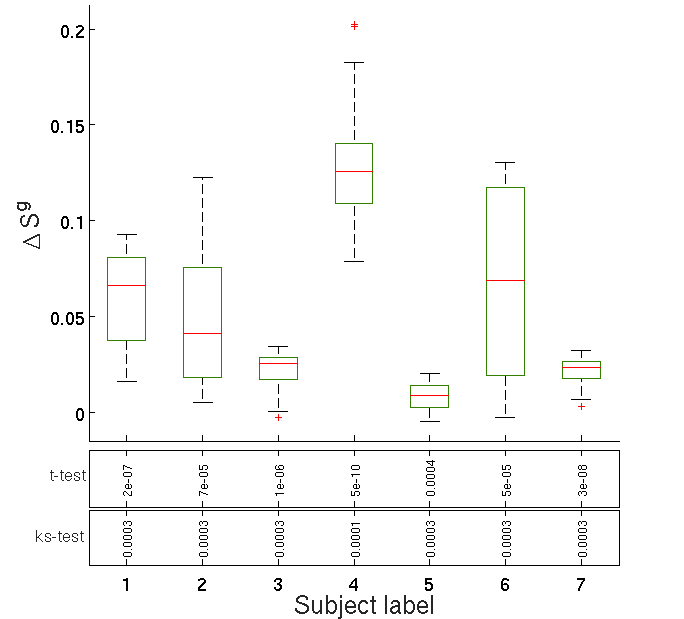}
   \caption{Nodal entropy before and after Ayahuasca. The 
boxplot shows the averaged geodesic entropy in 16 networks with the mean degree from 
$<k>=24$ to $<k>=39$ for each subject before (blue) and after (green) Ayahuasca. 
The median value increases for all of them.} %\textcolor{red}{figura fantastica!!!}}
  \label{boxplot}
  \end{figure}  

\section{Discussion  and Conclusion}
The (often non-trivial) rules of interactions among the nodes of a
network determine the nature of its emergent behaviors.
In many cases, the network interactions are defined by the relative
position of each node in the network structure.
The role of a node in a network depends on how it is contextualized
inside the network.
In a highly connected network, a node does not interact only with its
first neighbors, but also interact indirectly with the other nodes.
The geodesic entropy quantifies the statistics (i,e. the the entropy
functional of the probability distribution) of the geodesic distances
from a given node to all other nodes in the network, by classifying
all nodes according to their neighborhood radii.

In summary, we evaluate the geodesic entropy of functional brain
network of subjects in the resting state before and after the
ingestion of the psychedelic brew Ayahuasca.
We find that nodes of the functional network during Ayahuasca
experience tends to have greater geodesic entropy than in the ordinary
condition, resulting in networks with higher characteristic geodesic
entropy.
Hence, the geodesic distances between nodes become less constrained on
average, i.e. their distribution becomes ``wider.''  
In a previous work, we showed that the entropy of the degree
distribution of brain functional connectivity networks under the
influence of Ayahuasca is greather than in the ordinary state
\cite{VIO17a}.
The entropy of degree distribution is a global measurement and networks with different patters can share the same degree distribution. 
The result presented in this paper suggests that the patterns can be 
less restricted under Ayahuasca influence than in ordinary condition and
it does not depend on the degree distributions.
The diversity of geodesic distances are more well-distributed contributing to
the flexibility of interaction of the networks. 

The hypothesis of entropy increases in some aspect of brain in
psychedelic states has been discussed in the literature
\cite{CAR14,PAP16,CAR18}.  This entropic brain hypothesis predicts
that the psychedelics state is associated with greather entropy
compared to the ordinary state. The hypothesis could explain the
increased flexibility in thoughts, facility to access suppressed
memory, increase of creativity, among others \cite{CAR14}.

In conclusion, we have shown how the geodesic entropy quantifies
locally the connectivity to the network globally.
Further, we have used entropy-degree diagrams to evaluate the role of
each node in the network, giving a clearer view of the network
topology and global connectivity.
The application to fMRI-based functional connectivity networks sheds
insights on how the brain changes under the influence of external
influences. In this study, we used Ayahuasca, but there is no reason
why the method could not be applied to a variety of drugs or
meditative states, etc.
We hope that these ideas and methods find use in furthering our
understanding of complex networks in general and in brain function
networks specifically.

%
% 
% \paragraph{The options of the cite command itself}
% Please note that optional arguments to the \emph{key} change the reference in the bibliography, 
% not the citation in the body of the document. 
% For the latter, use the optional arguments of the \verb+\cite+ command itself:
% \verb+\cite+ \texttt{*}\allowbreak
% \texttt{[}\emph{pre-cite}\texttt{]}\allowbreak
% \texttt{[}\emph{post-cite}\texttt{]}\allowbreak
% \verb+{+\emph{key-list}\verb+}+.
\clearpage
%\newpage
\onecolumngrid
{\color{blue} 
\section{Supplementary information}
}
 %\subsection{sk- diagram and nodal entropy for all subjects}
\begin{figure}[h!]
\begin{center}
\vspace{-1.9 cm}
\hspace{1.5 cm} \hspace{-0.25cm}\includegraphics[width=0.02\linewidth,angle=-90]{before.png} 
\hspace{1 cm}\hspace{-0.25cm}\includegraphics[width=0.02\linewidth,angle=-90]{after.png} 

 \vspace{-.3 cm}
\hspace{-0.1cm}\includegraphics[width=0.2\linewidth]{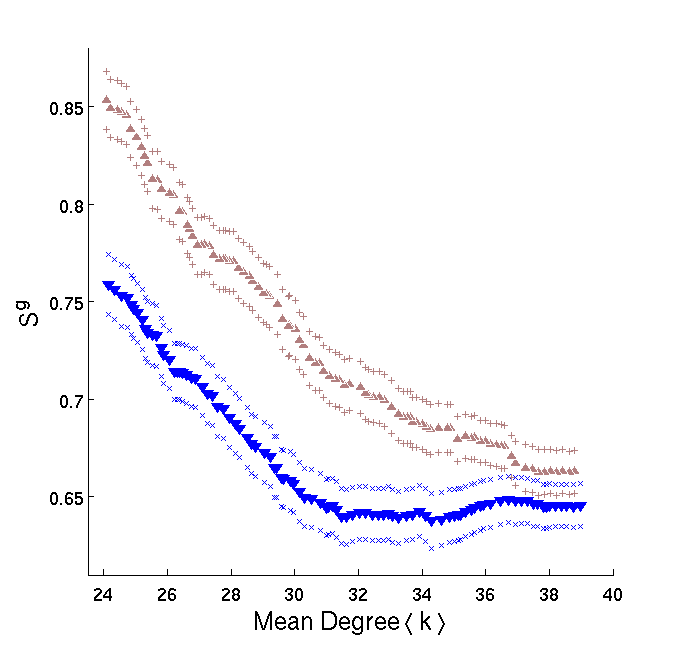} 
\vspace{-.3 cm}
\includegraphics[width=0.2\linewidth]{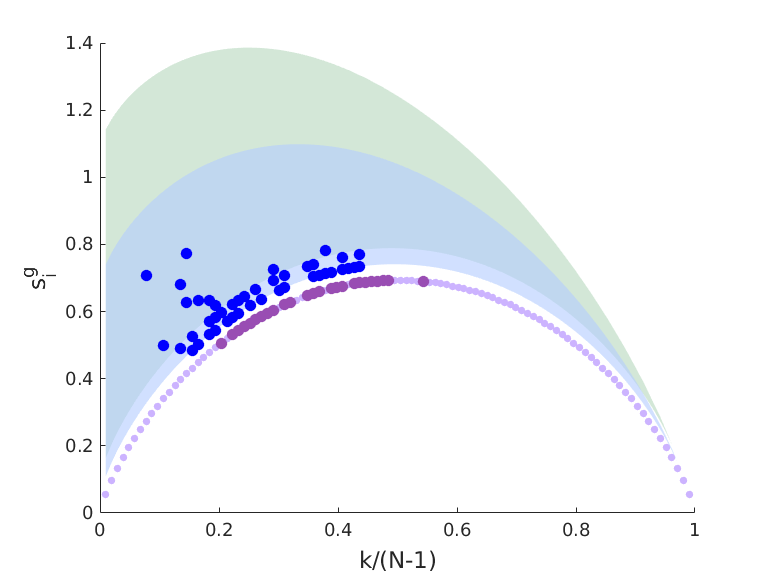} 
\vspace{-.3 cm}
\includegraphics[width=0.2\linewidth]{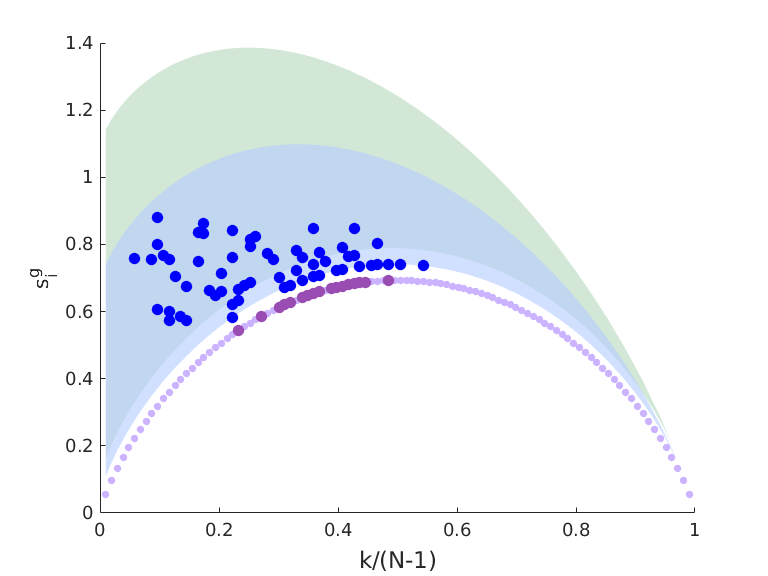} 

\hspace{-0.1cm}\includegraphics[width=0.2\linewidth]{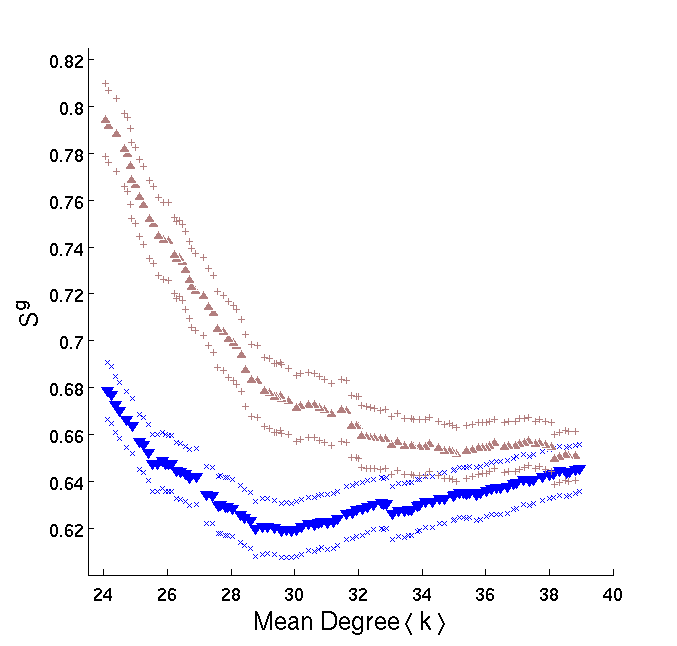} 
\includegraphics[width=0.2\linewidth]{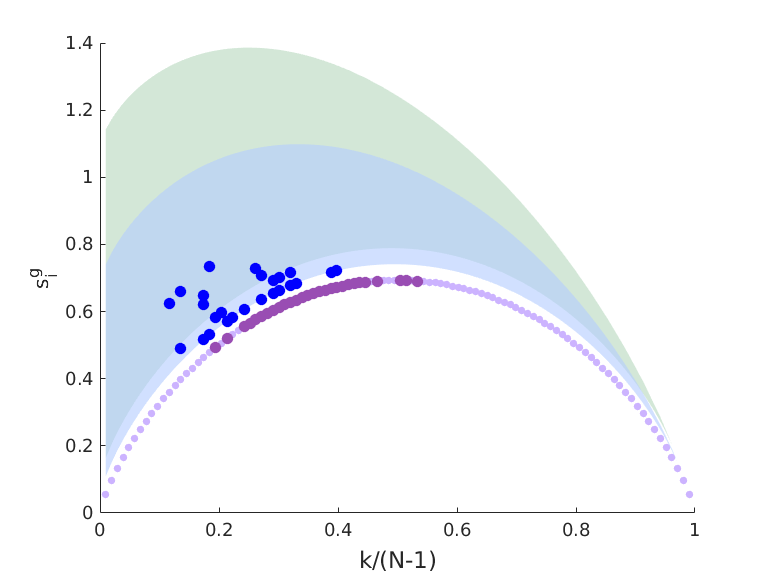} 
\includegraphics[width=0.2\linewidth]{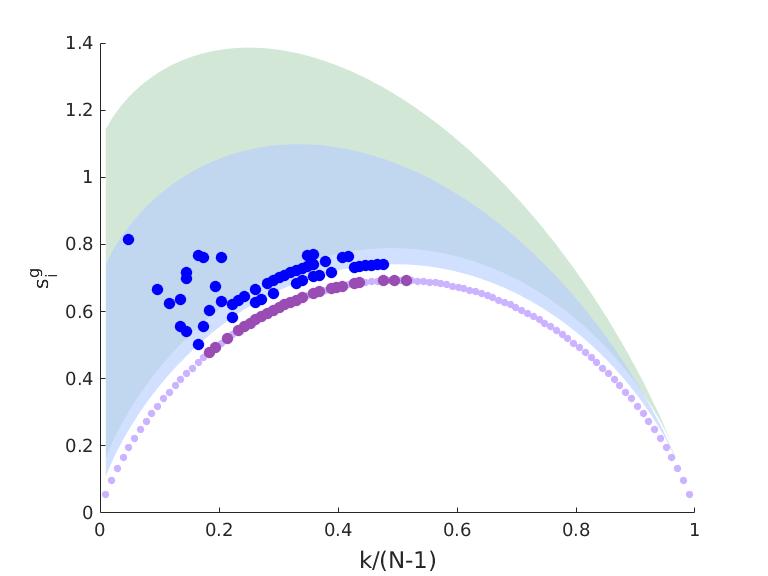} 

\hspace{-0.1cm}\includegraphics[width=0.2\linewidth]{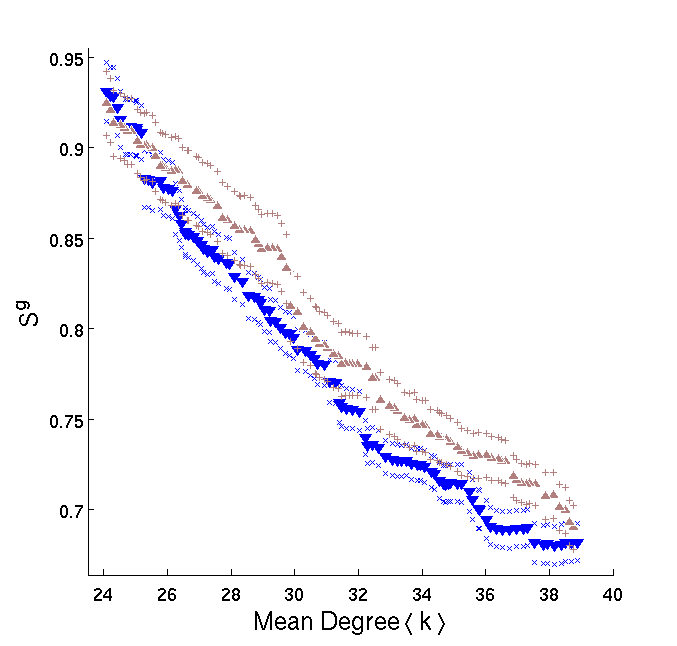} 
\includegraphics[width=0.2\linewidth]{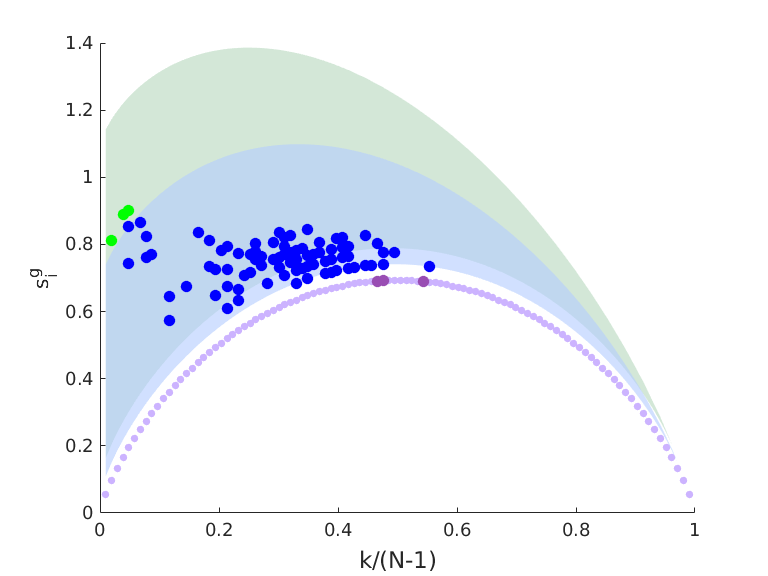} 
\includegraphics[width=0.2\linewidth]{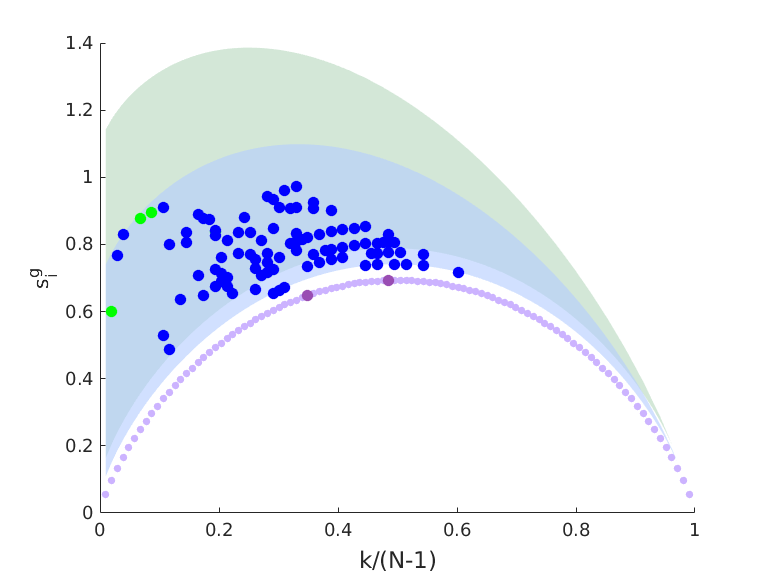} 

\hspace{-0.1cm}\includegraphics[width=0.2\linewidth]{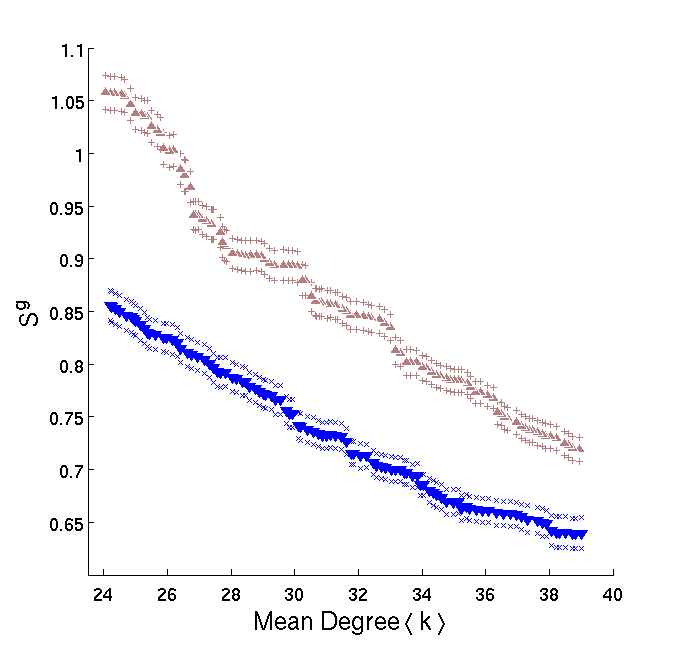}
\includegraphics[width=0.2\linewidth]{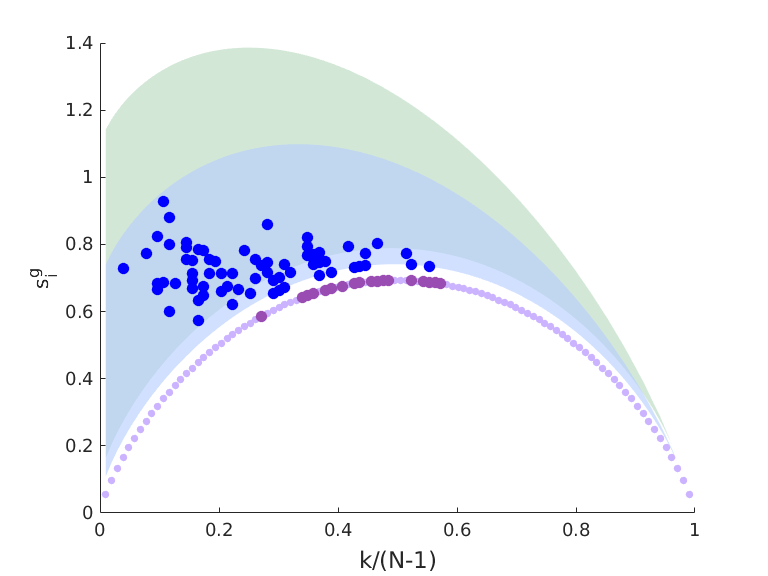} 
\includegraphics[width=0.2\linewidth]{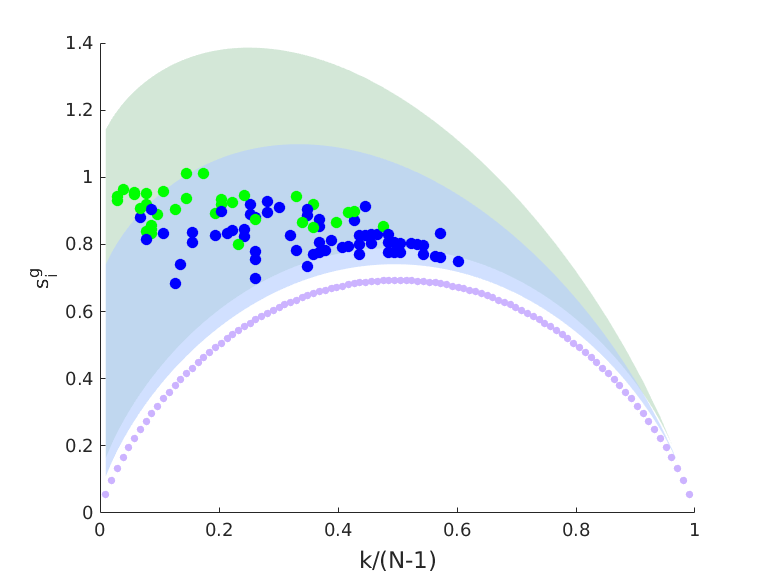} 

\hspace{-0.1cm}\includegraphics[width=0.2\linewidth]{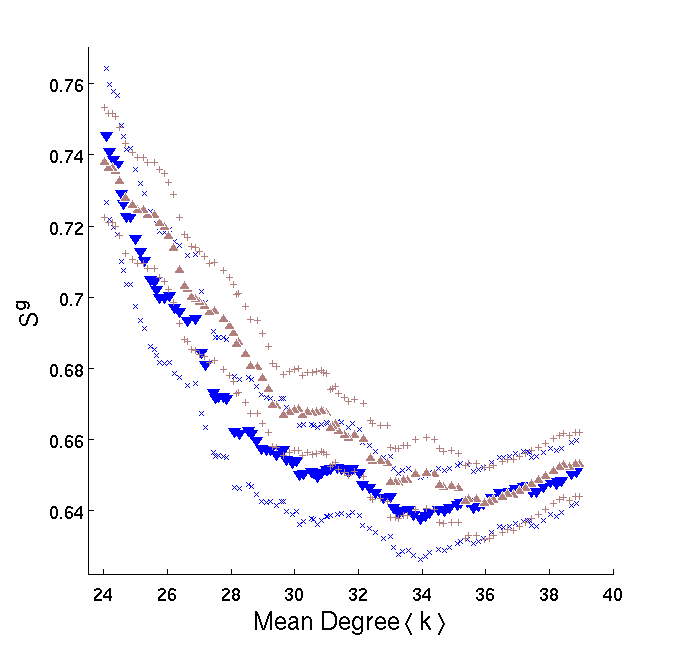} 
\includegraphics[width=0.2\linewidth]{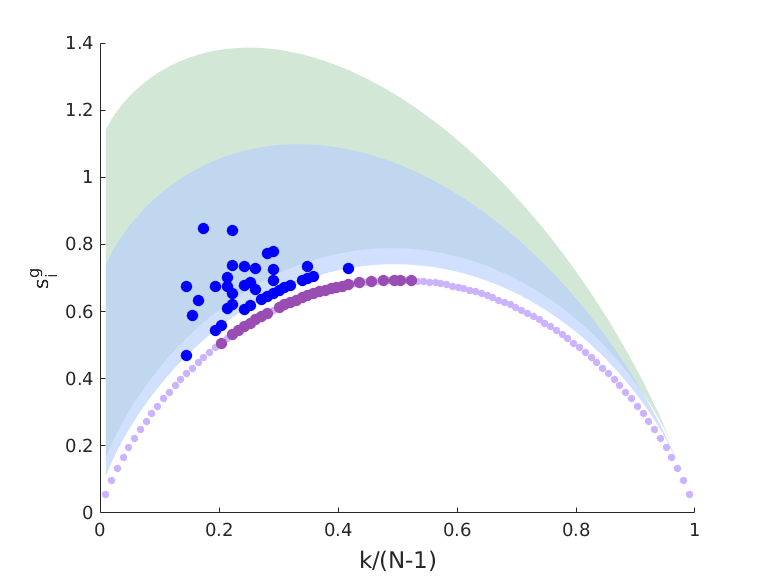} 
\includegraphics[width=0.2\linewidth]{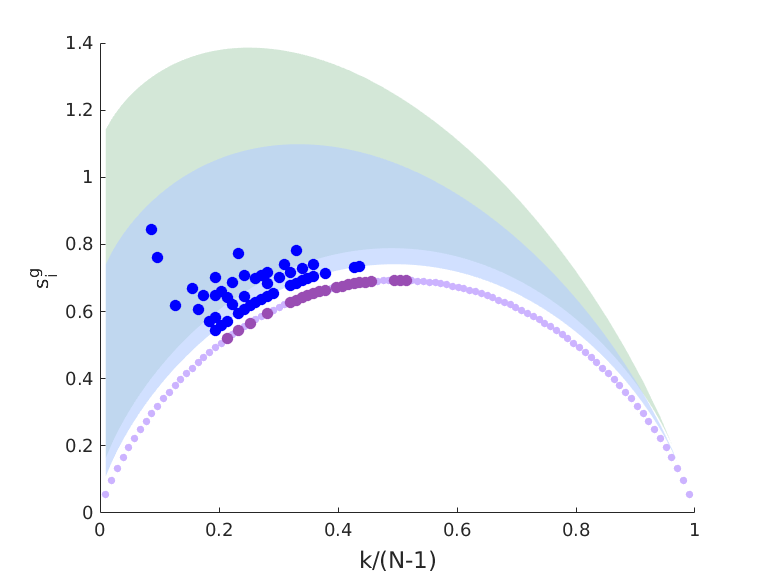} 

\hspace{-0.1cm}\includegraphics[width=0.2\linewidth]{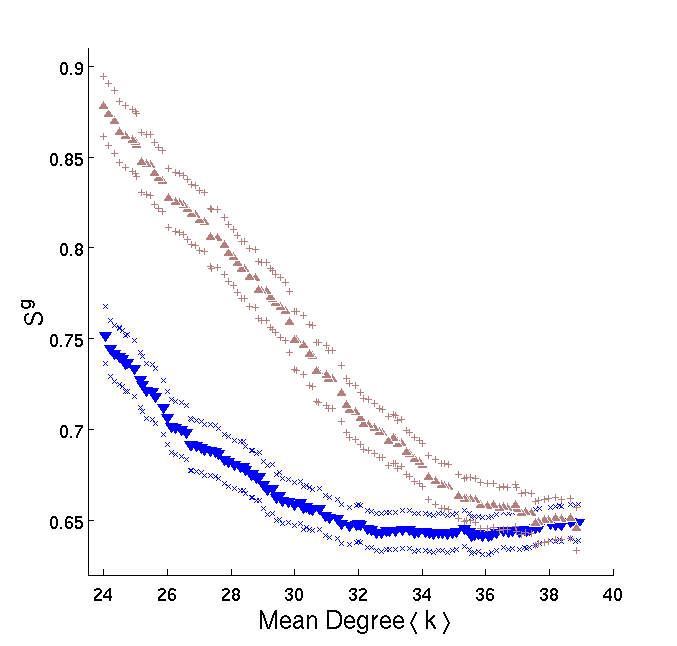} 
\includegraphics[width=0.2\linewidth]{s8_run1_32_norm_set.png} 
\includegraphics[width=0.2\linewidth]{s8_run2_32_norm_set.png}

\hspace{-0.11cm}\includegraphics[width=0.2\linewidth]{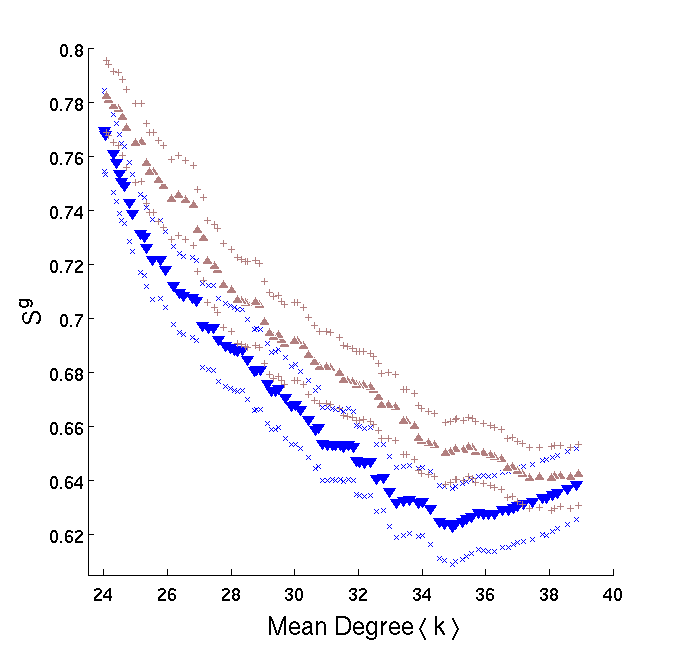} 
\includegraphics[width=0.2\linewidth]{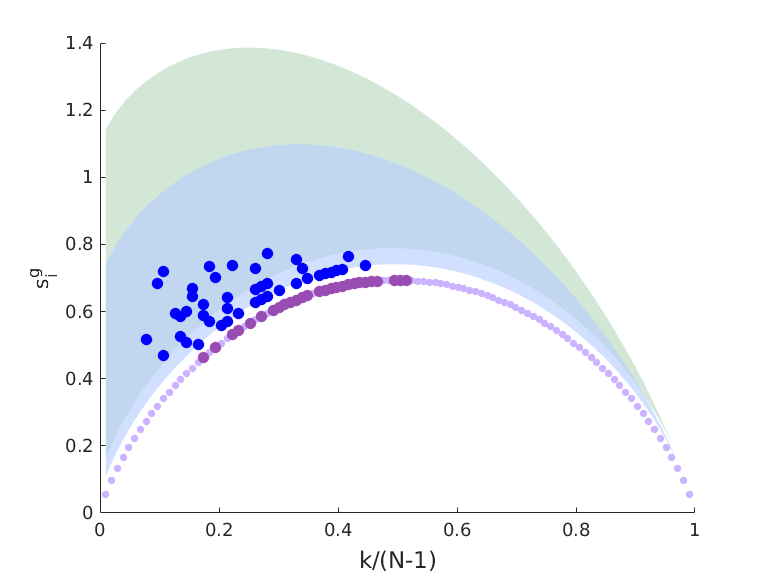} 
\includegraphics[width=0.2\linewidth]{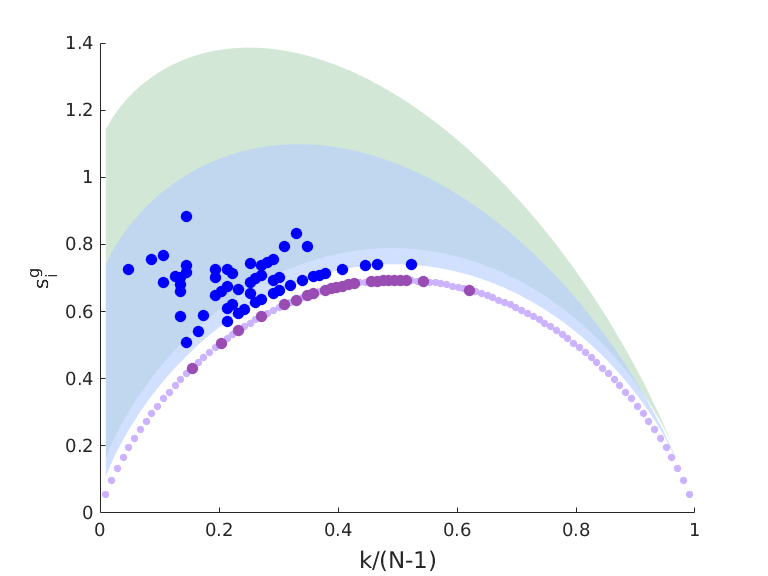} 

\label{diagrams}

\caption{Geodesic entropy before and after Ayahuasca intake for all subjects.
The first column depicts the curves to characteristic geodesic entropy for networks
with the mean degree from $\langle k \rangle=24$ to $\langle k \rangle=39$ for before (blue)
and after (brown) Ayahuasca intake. The second and third columns show the sk-diagram for
before and after respectively ( networks with mean degree $\langle k \rangle=32$ ). 
Note an increase of geodesic entropy for all subjects. 
}
\end{center}
\end{figure}

\bibliographystyle{apsrev4-1}

\clearpage

%\section*{References}

\bibliography{ne}

\clearpage

\end{document}